\documentclass[aps,twocolumn,superscriptaddress,floatfix,prc,twoside,tightenlines,nofootinbib,showpacs]{revtex4}
\usepackage[sort&compress]{natbib}
\usepackage{graphicx,mathrsfs,amsmath,amssymb,amsopn,mathbbol,bm}
\usepackage{color}

\begin{document}

\newcommand{\eg}{{\it e.g.}}
\newcommand{\etal}{{\it et. al.}}
\newcommand{\ie}{{\it i.e.}}
\newcommand{\be}{\begin{equation}}
\newcommand{\ee}{\end{equation}}
\newcommand{\bea}{\begin{eqnarray}}
\newcommand{\eea}{\end{eqnarray}}
\newcommand{\bfig}{\begin{figure}}
\newcommand{\efig}{\end{figure}}
\newcommand{\bce}{\begin{center}}
\newcommand{\ece}{\end{center}}

\title{Production of Light Nuclei at Thermal Freezeout in Heavy-Ion Collisions}
\author{Xinyuan Xu}
\affiliation{Cyclotron Institute and Department of Physics~\&~Astronomy, Texas A\&M
University, College Station, Texas 77843-3366, U.S.A.}
\affiliation{Department of Physics, University of Science and Technology of China, Hefei,
Anhui 230026, China}
\author{Ralf Rapp}
\affiliation{Cyclotron Institute and Department of Physics~\&~Astronomy, Texas A\&M
University, College Station, Texas 77843-3366, U.S.A.}

\date{\today}

\begin{abstract}
We revisit the problem of the production of light atomic nuclei in ultrarelativistic heavy-ion
collisions. While their production systematics is well produced by hadro-chemical freezeout
at temperatures near the QCD pseudo-critical temperature, their small binding energies of
a few MeV per nucleon suggest that they cannot survive as bound states under these conditions. 
Here, we adopt the concept of effective chemical potentials in the hadronic evolution from
chemical to thermal freezeout (at typically $T_{\rm fo}$$\simeq$100\,MeV), which, despite
frequent elastic rescatterings in hadronic matter, conserves the effective numbers of 
particles which are stable under strong interactions, most notably pions, kaons and nucleons. 
It turns out that the large chemical potentials that build up for antibaryons result in
thermal abundances of light nuclei and antinuclei, formed at thermal freezeout, which essentially 
agree with the ones evaluated at chemical freezeout. Together with their transverse-momentum 
spectra, which 
also indicate a kinetic freezeout near $T_{\rm fo}$, this provides a natural explanation for 
their production systematics without postulating their survival at high temperatures.   
\end{abstract}

\maketitle

\section{Introduction}
\label{sec:intro}
The main goal of ultrarelativistic heavy-ion collisions (URHICs) is, broadly speaking, the 
production and characterization of hot and dense QCD matter, thereby recreating conditions 
akin to the first few microseconds in the evolution of our Universe. The current picture of 
the time evolution of a central URHIC is as follows: a short pre-equilibrium phase (of duration 
$\Delta \tau_{\rm pre} \lesssim 0.5$\,fm/$c$) is followed by hydrodynamically expanding 
quark-gluon plasma (QGP, $\Delta \tau_{\rm QGP} \simeq 5$\,fm/$c$), hadronization with a 
chemical freezeout of the stable hadron species at $T_{\rm H}\simeq T_{\rm ch}\simeq 160$\,MeV, 
followed by a hadronic phase ($\Delta \tau_{\rm had} \simeq 5-10$\,fm/$c$) with strong elastic 
collisions (mostly resonance scattering) which maintain approximate local thermal equilibrium 
until a freezeout temperature of about $T_{\rm fo}\simeq 100$\,MeV. By that time the fireball
has acquired average expansion velocities in excess of half the speed of light, as deduced from 
systematic analyses of transverse-momentum spectra of pions, kaons and 
protons~\cite{Abelev:2008ab,Abelev:2013vea}.        
In this context, the production of light nuclei and antinuclei has received considerable 
attention in recent years, see Refs.~\cite{Bellini:2018epz,Chen:2018tnh} for a recent reviews. 
On the one hand, 
URHICs have enabled the first man-made production of antinuclei beyond nuclear number A=3. On 
the other hand, it was found that the production ratios of the observed nuclei, which by now 
encompass anti-/deuterons, anti-/$^3$He and anti-/$^4$He (as well as light hyper anti-/nuclei), 
closely follow the predictions of the statistical hadronization model at chemical 
freezeout~\cite{Andronic:2010qu,Andronic:2016nof}. At the same time, the $P_T$ spectra of the
nuclei indicate a kinetic decoupling close to thermal freezeout, suggesting that they participate 
in the elastic interactions of the hadronic phase. This, however, poses a puzzle, as it is 
rather surprising that composite objects with binding energies that are about an order
of magnitude smaller than the ambient temperature can survive in this environment (cf.~also
Ref.~\cite{Andronic:2017pug} for a recent discussion of this issue).   
As an alternative, coalescence models have been employed to form the nuclei in the late stages 
of the evolution, near thermal freezeout~\cite{Kapusta:1980zz,Scheibl:1998tk}. While these 
models can account for finite-size (or quantum) effects in the formation process, they also
require extra parameters, \eg, the so-called coalescence parameter $B_A$ (fixing the overall 
normalization), and often involve an assumption of near-collinearity of the coalescing nucleon 
momenta (which, strictly speaking, violates energy conservation). 

In the present paper we investigate another option which does not involve additional parameters
relative to the chemical freezeout point as specified by a temperature ($T_{\rm H}$  and 
pertinent baryon chemical potential ($\mu_B^{\rm H}$). We will adopt the concept of partial 
chemical equilibrium throughout the hadronic evolution from $T_{\rm H}$ to $T_{\rm fo}$ by 
introducing chemical potentials for the hadrons which are stable under strong 
interactions~\cite{Bebie:1992,Hung:1998,Rapp:2002fc,Teaney:2002aj,Hirano:2002ds}. This essentially 
ensures that the observed ratios of strongly stable hadrons (including strong feeddown contributions) 
are conserved in the elastic interactions which maintain local thermal equilibrium throughout the 
hadronic phase. An important variation of this theme, critical for the present analysis, is the 
separate conservation of baryon ($B$) and antibaryon ($\bar B$) number~\cite{Rapp:2002fc}. The large 
(strong-interaction) annihilation cross sections of baryons and antibaryons into multi-meson states, 
\eg, $N\bar N \to 5\pi$, naively suggest a decline in the total number of baryons plus antibaryons. 
However, in the presence of large meson-chemical potentials the backward reaction, \eg, 
$5\pi\to N\bar N $, as dictated by detailed balance, can largely uphold the $\bar B/B$ ratios 
created at chemical freezeout~\cite{Rapp:2000gy}.
With the such obtained baryon and antibaryon chemical potentials~\cite{Rapp:2002fc} we 
predict the production ratios of the light nuclei, and compute their transverse-momentum ($P_T$) 
spectra, at thermal freezeout, which mitigates the issue of their survival at high temperatures. 
 
The remainder of this article is organized as follows. In Sec.~\ref{sec:chem} we recall the basic 
elements of partial chemical equilibrium and the construction of the effective chemical potential
for mesons and anti-/baryons in the hadronic phase of URHICs. In Sec.~\ref{sec:ratios} we present
our main results for the production ratios of light anti-/nuclei evaluated using the predicted
chemical potentials at {\em kinetic} freezeout. In Sec.~\ref{sec:pt} we evaluate their 
$P_T$ spectra at the same kinetic freezeout. In Sec.~\ref{sec:concl} we conclude.  

\section{Chemical Freezeout and Potentials}
\label{sec:chem}
The concept of hadro-chemical freezeout in the fireball evolution of a heavy-ion collision
implies that from this point on the observed hadron production ratios (or yields, upon specifying 
a single 3-volume, $V_{\rm FB}(T_{\rm ch})$, to match the total yields to experiment at a given 
centrality) do no longer change. This concept has been very successful in the description of
the hadron production yields as observed in 
experiment~\cite{BraunMunzinger:2003zd,Cleymans:2005xv,Becattini:2005xt,Andronic:2008gu}. 
Typical chemical-freezeout temperatures at ultrarelativistic collision energies are close to 
the pseudo-critical temperature extracted from lattice QCD (lQCD) at vanishing baryon chemical 
potential, $T_{\rm ch}\simeq160\,{\rm MeV}\simeq T_{\rm pc}$, while the pertinent baryon chemical 
potential, $\mu_B^{\rm ch}$, varies from $\sim$240\,MeV at top SPS energy ($\sqrt{s}$=17.3\,GeV) 
via $\sim$25\,MeV at top RHIC energy ($\sqrt{s}$=200\,GeV) to $\lesssim$1\,MeV at LHC energies 
($\sqrt{s}$ of several TeV).
The microscopic origin of chemical freezeout is associated with the smallness of the cross section 
of chemistry-changing inelastic interactions of typically a few mb (\eg, for $\pi\pi\to K\bar{K}$). 
On the other hand, quasi-elastic resonant interactions, such as $\pi\pi\to\rho\to\pi\pi$ or 
$\pi N \to\Delta\to \pi N$ are up to 2 orders of magnitude larger and thus maintain local thermal
equilibrium in the expanding fireball down to kinetic freezeout temperatures of 
$T_{\rm fo}\simeq100$\,MeV. These reactions ``conserve" the chemistry (total abundances) of the 
stable particle (such as pions, kaons, etas, nucleons, etc.) which can be accounted for by effective 
chemical potentials ($\mu_{\pi,K,\eta,N,...}$). Within this scheme, the strong resonances, such as 
$\rho$ and $\Delta$, which are essential for the thermodynamics of the hadronic medium, acquire 
effective chemical potentials according to the law of mass action, \eg, $\mu_\Delta = \mu_\pi +\mu_N$ 
and $\mu_\rho=2\mu_\pi$. With decreasing temperature, substantial positive meson chemical potentials 
build up, compensating the suppression of their numbers relative 
to the conserved baryon number. They typically reach near 100(150)\,MeV for pions (kaons) at kinetic 
freezeout under both RHIC and LHC conditions, cf.~Fig.~\ref{fig_chem} (the RHIC results shown here
differ from the ones originally shown in Ref.~\cite{Rapp:2002fc} by updating the chemical freezeout
temperature from $T_{\rm ch}$=180\,MeV to 160\,MeV). 

\bfig[!t]
\begin{minipage}{0.5\textwidth}
\includegraphics[width=0.98\textwidth]{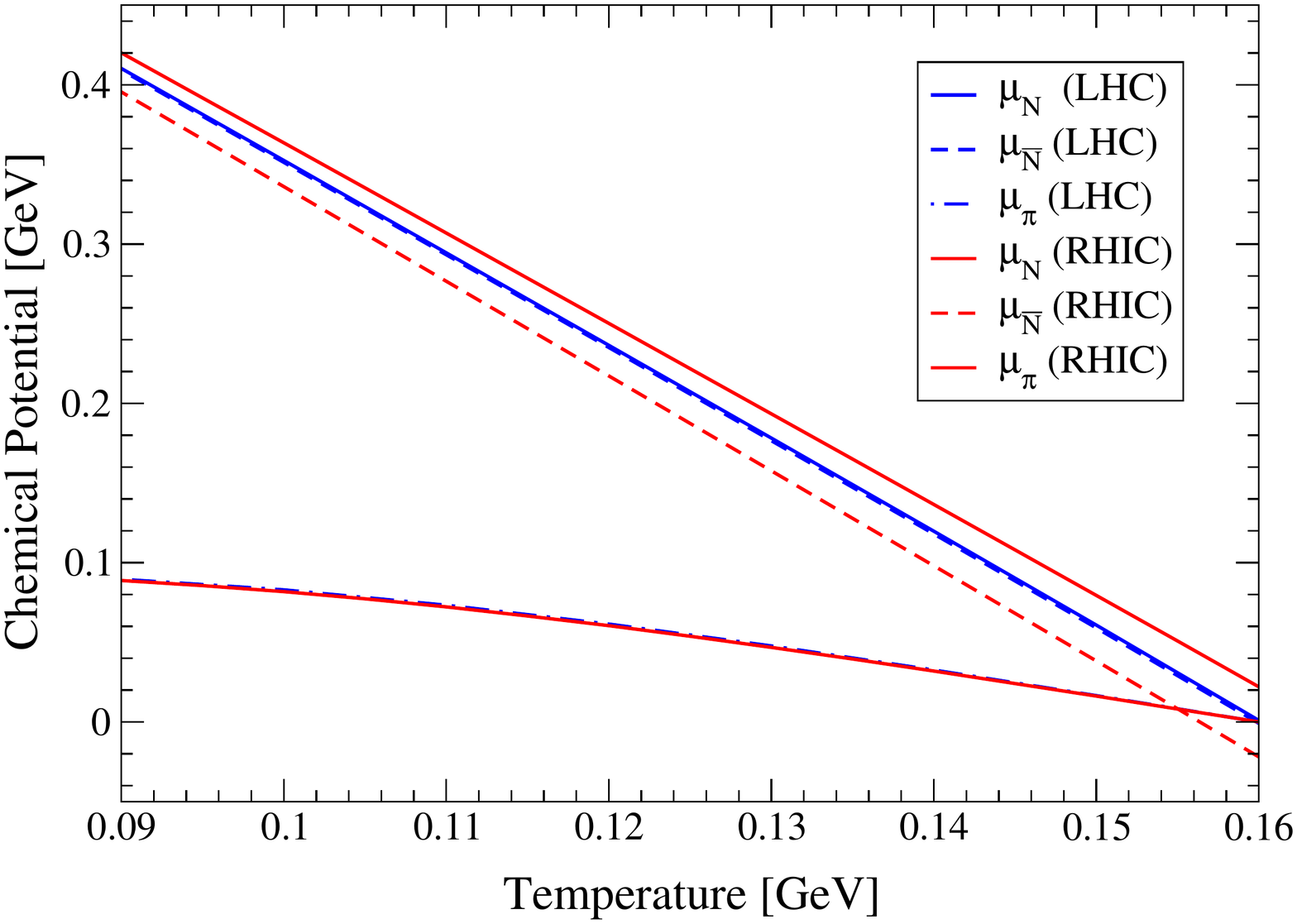}
\end{minipage}
\caption{Temperature dependence of chemical potentials for nucleons (solid lines),
antinucleons (dashed lines) and pions (dash-dotted lines) at RHIC (red curves) and the 
LHC (blue curves). The curves for the nucleon and antinucleon chemical potentials at the 
LHC (solid and dashed blue, respectively), as well as for the pion chemical potentials 
at RHIC and the LHC (dash-dotted red and blue, respectively), almost coincide.}
\label{fig_chem}
\efig

An exception to the small inelastic hadronic cross sections is the annihilation of baryons and 
antibaryons into multi-meson states, with cross sections of up to 100\,mb in the low-energy range 
relevant for a thermal heat bath at temperatures of $T\simeq150$\,MeV. This suggests substantial 
annihilation rates of antibaryons 
in the hadronic phase of the fireball and has raised the question of how to reconcile the observed 
yields of, \eg, antiprotons with the chemical-freezeout picture. In Ref.~\cite{Rapp:2000gy} it was 
shown that, in the presence of large meson chemical potentials, the inverse reaction of multi-meson 
fusion into $B\bar B$ largely compensates the $B\bar B$ annihilation reactions and thus stabilizes 
the antibaryon abundances throughout the hadronic evolution of the fireball. In essence, this amounts 
to yet another effective chemical potential, $\mu_{\bar B}^{\rm eff}$, that separately conserves the 
antibaryon number so that, \eg, $\mu_{\bar N} = -\mu_N + \mu_{\bar B}^{\rm eff}$, or 
$\mu_{\bar\Delta} = -\mu_{\Delta} + \mu_{\bar B}^{\rm eff}$. The temperature dependence of the
nucleon and antinucleon chemical potentials is also shown in Fig.~\ref{fig_chem} for both RHIC and 
LHC conditions. It illustrates the role of the effective antibaryon chemical potential, which reaches 
close to 800\,MeV at the lowest temperatures and results in a near constant antiproton-to-proton ratio, 
$\bar{p}/p \simeq \exp[-(\mu_N-\mu_{\bar N})/T] \simeq \exp[-(2\mu_N+\mu_{\bar B}^{\rm eff})/T]$.
At RHIC, with $\mu_N^{\rm ch}\simeq25\,MeV$ at chemical freezeout, the nucleon and antinucleon
chemical potentials are notably different while at the LHC, where $\mu_N^{\rm ch}\simeq1\,MeV$, 
they are close to each other. All of them reach around 400\,MeV near thermal freezeout. The baryon
and antibaryon numbers are separately conserved, and thus their sum as well. This has important
phenomenological implications for, \eg, low-mass dilepton spectra where the medium effects on the
the $\rho$ (and $\omega$) meson in the hadronic phase are equally driven by interactions with 
baryons and antibaryons, and are thus sensitive to the total sum of their densities (see, \eg, 
Ref.~\cite{Rapp:2013nxa} for a recent review). 

\section{Production Ratios of Anti-/Nuclei}
\label{sec:ratios}
We are now in position to apply the hadro-chemistry described in the previous section to
production ratios of light anti-/nuclei as recently measured in URHICs. In Tab.~\ref{tab_nuclei}
we summarize the masses ($m_X$) and spin degeneracies ($d_X$) of the pertinent particles. 
In the following, we simply compute the densities of these particles as a function
of temperature and chemical potential, 
\begin{equation} 
n_X(\mu_X,T) = d_X  \int \frac{d^3P}{(2\pi)^3} f^X(E_X;\mu_X,T) \ , 
\end{equation}
where $E_X=\sqrt{m_X^2 + P^2}$ is their energy,  $\mu_X={\rm A} \mu_N$ for nucleus $X$,
$\mu_{\bar{X}} = {\rm A} (-\mu_N +\mu_B^{\rm eff})$ for its antinucleus (A: nuclear mass number),
and $f^X=\exp(-(E_X-\mu_X)/T)$ is the Boltzmann distribution (quantum-statistical corrections
can be safely neglected). Neglecting feeddown contributions, we then take suitable ratios
of these densities or evaluate total yields by specifying a single 3-volume at given centrality.  
\begin{table}[!t]
\label{tab_nuclei}
\begin{center}
\begin{tabular}{|c|c|c|}
\hline
 (Anti-) Nucleus & $m_X$ [MeV] & $d_X$ 
\\
\hline  \hline
 $d$ ($\bar{d}$)  & 1875.6  &  3    
\\
\hline 
$t$ ($\bar{t}$)   &   2808.9 & 2 
\\
\hline
$^3$He ($^3\bar{\rm He}$) & 2809.3 & 2
\\
\hline
$^4$He ($^4\bar{\rm He}$) & 3728.3& 1 
\\
\hline
\end{tabular}
\end{center}
\caption{The masses (second column) and degeneracies (third column) of the light anti-/nuclei 
(first column) considered in this work.}
\end{table}

\bfig[!t]

\vspace{-0.5cm}

\begin{minipage}{0.5\textwidth}
\includegraphics[width=0.95\textwidth]{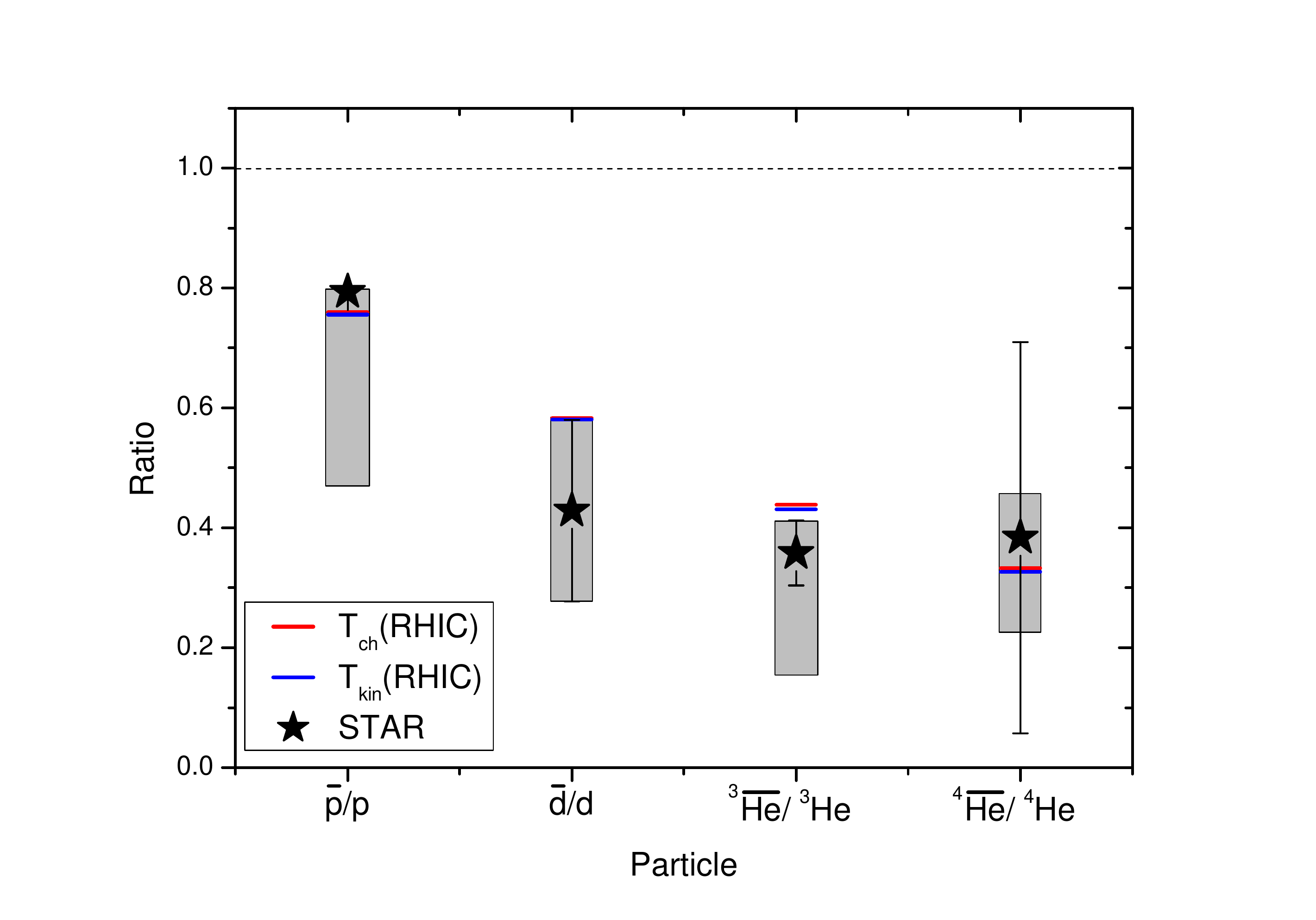}
\end{minipage}

\vspace{-0.5cm}

\begin{minipage}{0.5\textwidth}
\includegraphics[width=0.95\textwidth]{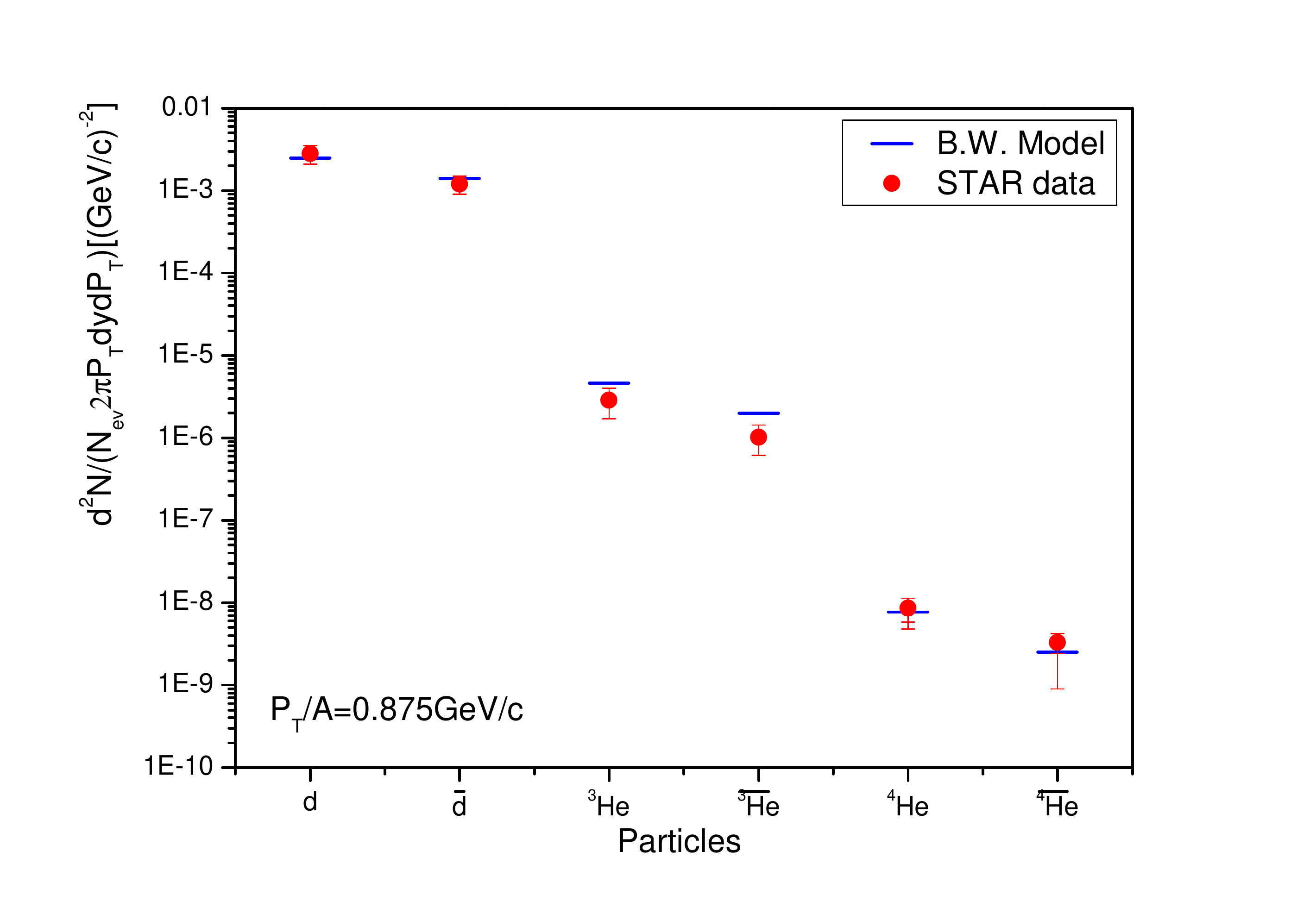}
\end{minipage}
\caption{Production ratios (upper panel) and $P_T$-differential invariant yields  
for $P_T/$A=0.875\,GeV (lower panel) of light anti-/nuclei in central Au+Au collisions 
at $\sqrt{s_{\rm NN}}$=0.2\,TeV at RHIC at chemical (red bars) and kinetic (blue bars) 
freezeout, compared to STAR data~\cite{Abelev:2009ae,Agakishiev:2011ib}.
}
\label{fig_rhic}
\efig
We start our discussion at RHIC energies. The upper panel of Figure~\ref{fig_rhic} shows 
the antiparticle-to-particle ratios, computed at both chemical and kinetic freezeout,
in comparsion to STAR data for anti-/deuterons, anti-/$^3$He and 
anti-/$^4$He~\cite{Abelev:2009ae,Agakishiev:2011ib}. The experimental ratios are obtained 
from invariant yields in $P_T$ bins of width 0.25\,GeV around $P_T/A=0.875$\,GeV. As discussed 
in Ref.~\cite{Agakishiev:2011ib}, from estimates using blastwave fits, these ratios differ 
from the $P_T$-integrated yield ratios by $\sim$1\% or so. 
As expected from the construction of the chemistry-conserving evolution of the hadronic
medium from chemical down to thermal freezeout, the calculated nucleus-to-antinucleus
ratios vary little throughout the hadronic phase and show fair agreement with the data. 
For a direct comparison of our calculations to the absolute values of the $P_T$-invariant 
yields, we utilize our  blastwave fits discussed in the following section, 
since larger variations are expected for particles with different masses when going 
from finite $P_T$ bins to $P_T$ integrated yields. In this case, we adjust one overall
normalization constant in the anti-/deuteron sector and then predict the yields for
anti-/$^3$He and anti-/$^4$He. While the data are reproduced for the latter, a factor of 
$\sim$2 discrepancy is observed for the former, see lower panel of Fig.~\ref{fig_rhic}. 
A very similar trend is observed when comparing our model predictions for the 
$P_T$-integrated yields to the recent experimental results of anti-/deuterons and the 
triton~\cite{Yu:2017bxv,Liu:2018qm}.
The origin of this discrepancy remains unclear at this point.  

\bfig[!t]

\vspace{-0.5cm}

\begin{minipage}{0.5\textwidth}
\includegraphics[width=0.95\textwidth]{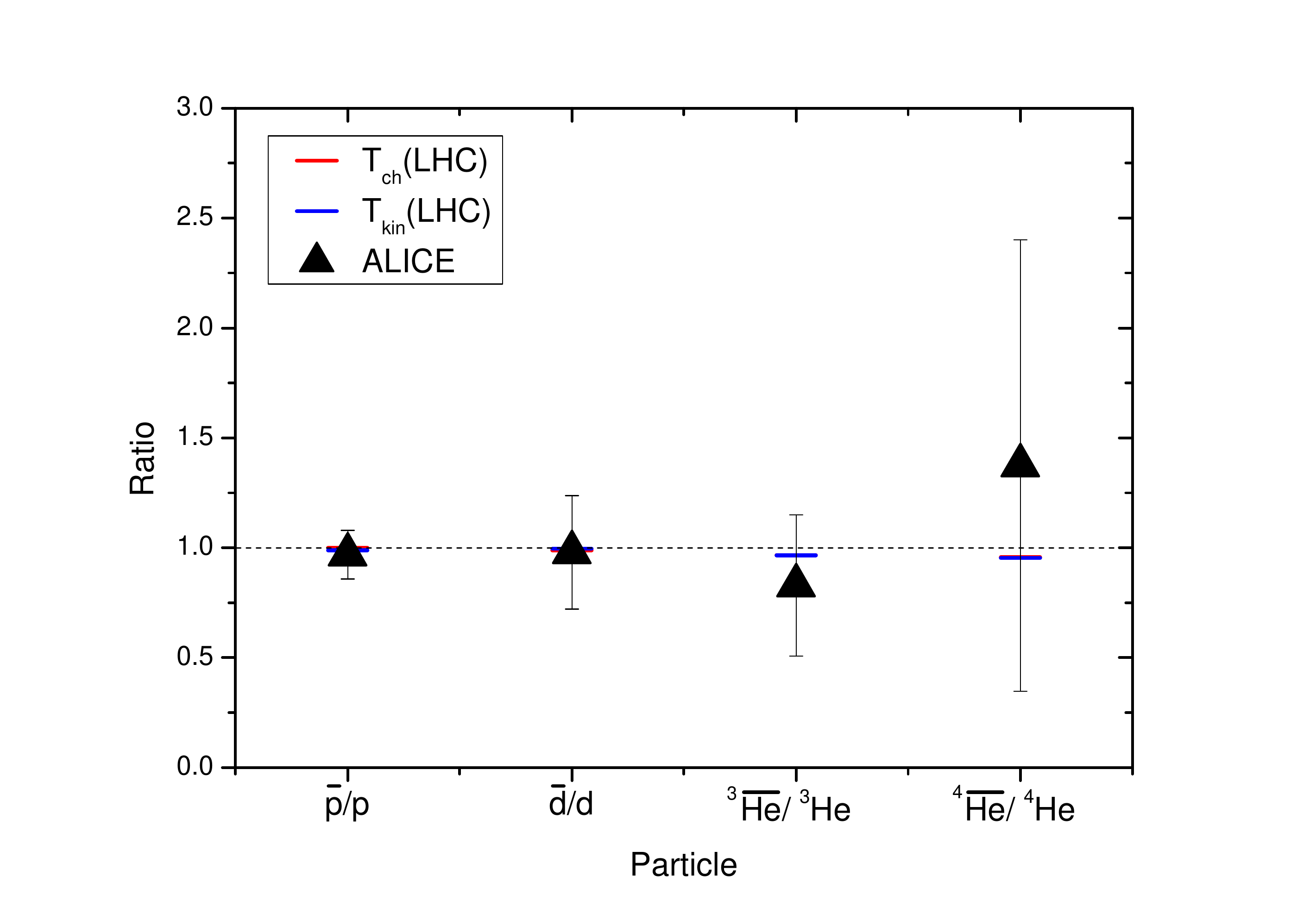}
\end{minipage}

\vspace{-0.5cm}

\begin{minipage}{0.5\textwidth}
\includegraphics[width=0.95\textwidth]{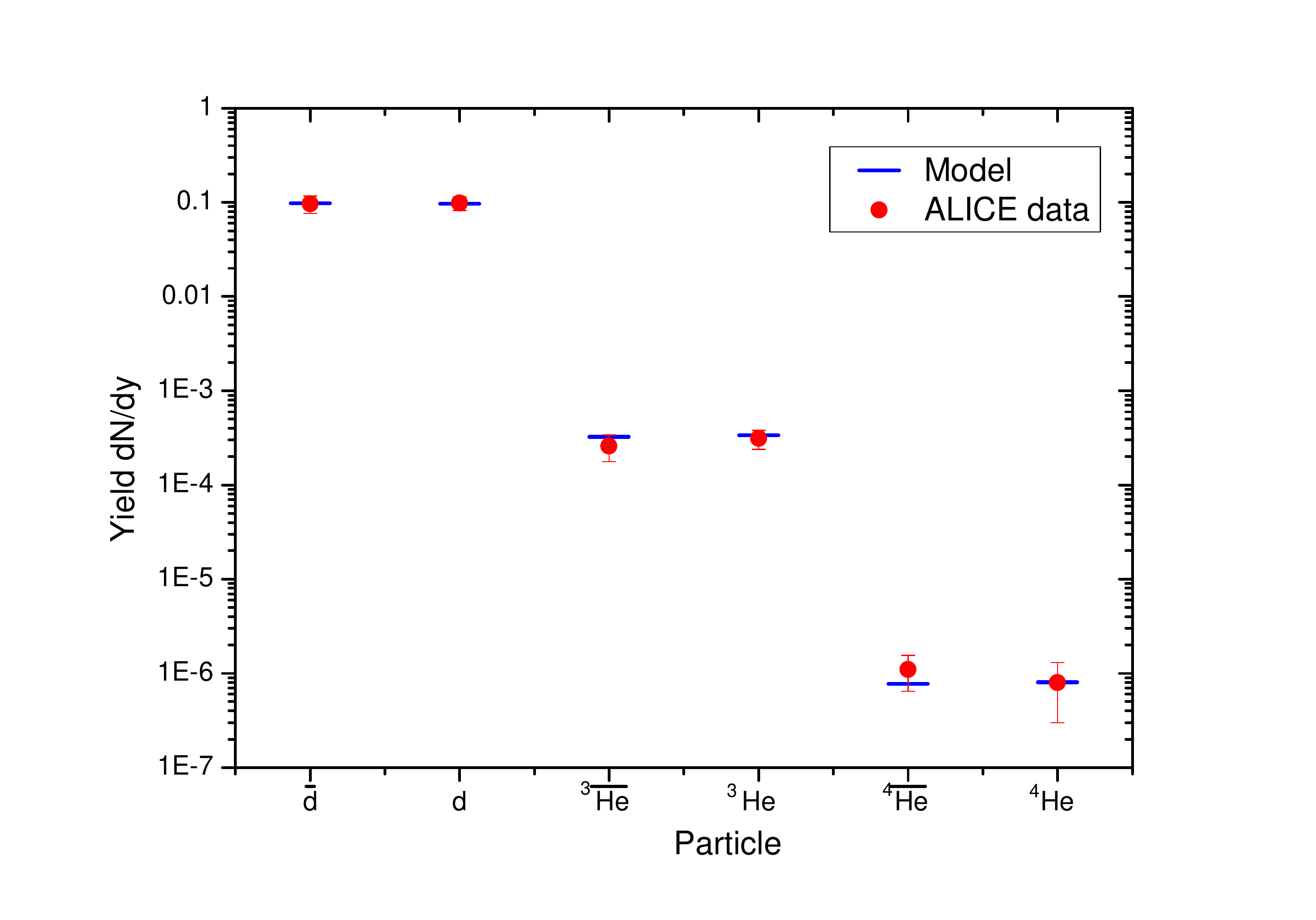}
\end{minipage}

\caption{Production ratios (upper panel) and absolute yields (lower panel) of light anti-/nuclei 
in 0-10\% central Pb+Pb collisions at $\sqrt{s_{\rm NN}}$=2.76\,TeV at the LHC at chemical 
(red bars) and kinetic (blue bars) freezeout, compared to ALICE 
data~\cite{Adam:2015vda,Acharya:2017bso}. 
}
\label{fig_lhc}
\efig
The results for antiparticle-to-particle ratios at the LHC are dislayed in Fig.~\ref{fig_lhc}
(upper panel).
Again, the calculated results differ little between chemical and kinetic freezeout, 
and agree with ALICE data within errors. The data for the ratios are compatible with one,
reflecting the near baryon-antibaryon symmetric (or net-baryon free) mid-rapidity region
at LHC energies. However, turning to the absolute yields (lower panel of Fig.~\ref{fig_lhc}), 
variations over about 6 orders of magnitude between anti-/deuterons and anti-/$^4$He are 
found, due to the exponential sensitivity to the particle mass. In our calculations, the 
absolute yields require to
fix one volume parameter for a given centrality, which we adjust to the central value
of the deuteron yield (we do not include excluded volume corrections here). The remaining
five yields for antideuterons, anti-/$^3$He and anti-/$^4$He are predictions, which show 
good agreement with the data. This, of course, has been noted before in statistical-model 
fits at chemical freezeout~\cite{Andronic:2010qu,Andronic:2016nof}; the new point here is 
that this description is upheld until kinetic freezeout, owing to the conserving chemistry 
in the hadronic evolution.

\section{Transverse-Momentum Spectra}
\label{sec:pt}
To further check the consistency of the description of the production mechanism of
light anti-/nuclei at  kinetic freezeout, we also revisit their transverse-momentum
spectra. Blastwave descriptions have been previously employed for nuclear $P_T$ spectra 
and indeed indicated temperatures and flow velocities  
near the thermal freezeout extracted from light-hadron ($\pi$, $K$, $p$) spectra. Here, we 
slightly modify this strategy by fixing the temperature to $T$=100\,MeV at both RHIC and 
the LHC and adjust the flow velocity and profile in a common fit to the different 
anti-/nuclei spectra. We employ the standard blastwave expression~\cite{Schnedermann:1993ws}, 
\begin{equation}
\frac{dN_X}{P_TdP_T} = C \int\limits_0^R rdr m_T I_0\left(\frac{P_T\sinh\rho}{T}\right)
K_1\left(\frac{m_T\cosh\rho}{T}\right) \ , 
\end{equation} 
with $m_T=\sqrt{m_X^2+P_T^2}$, a transverse flow rapidity $\rho=\tanh^-1(\beta_r)$ and 
flow profile $\beta_r=\beta_s\left(\frac{r}{R}\right)^n$ characterized by the exponent $n$. 
We find that with a surface flow velocity of $\beta_s$=0.780$c$  and $n$=1.2 for 
central Au-Au at RHIC,  as well as  $\beta_s$=0.866$c$ and $n$=0.8 for central Pb-Pb 
at the LHC, a fair description of the available $P_T$ spectra can be obtained, see 
Fig.~\ref{fig_pt}. The exception are again the A=3 nuclei at RHIC (as found for
the yields in the previous section). The larger flow at the LHC 
is expected due to the longer lifetime the fireball (which, at the same $T$, is 
larger than at RHIC), and is also in line with the trends extracted from fits to 
light-hadron spectra~\cite{Abelev:2008ab,Abelev:2013vea}.

\bfig[!t]

\vspace{-0.5cm}

\begin{minipage}{0.5\textwidth}
\includegraphics[width=0.95\textwidth]{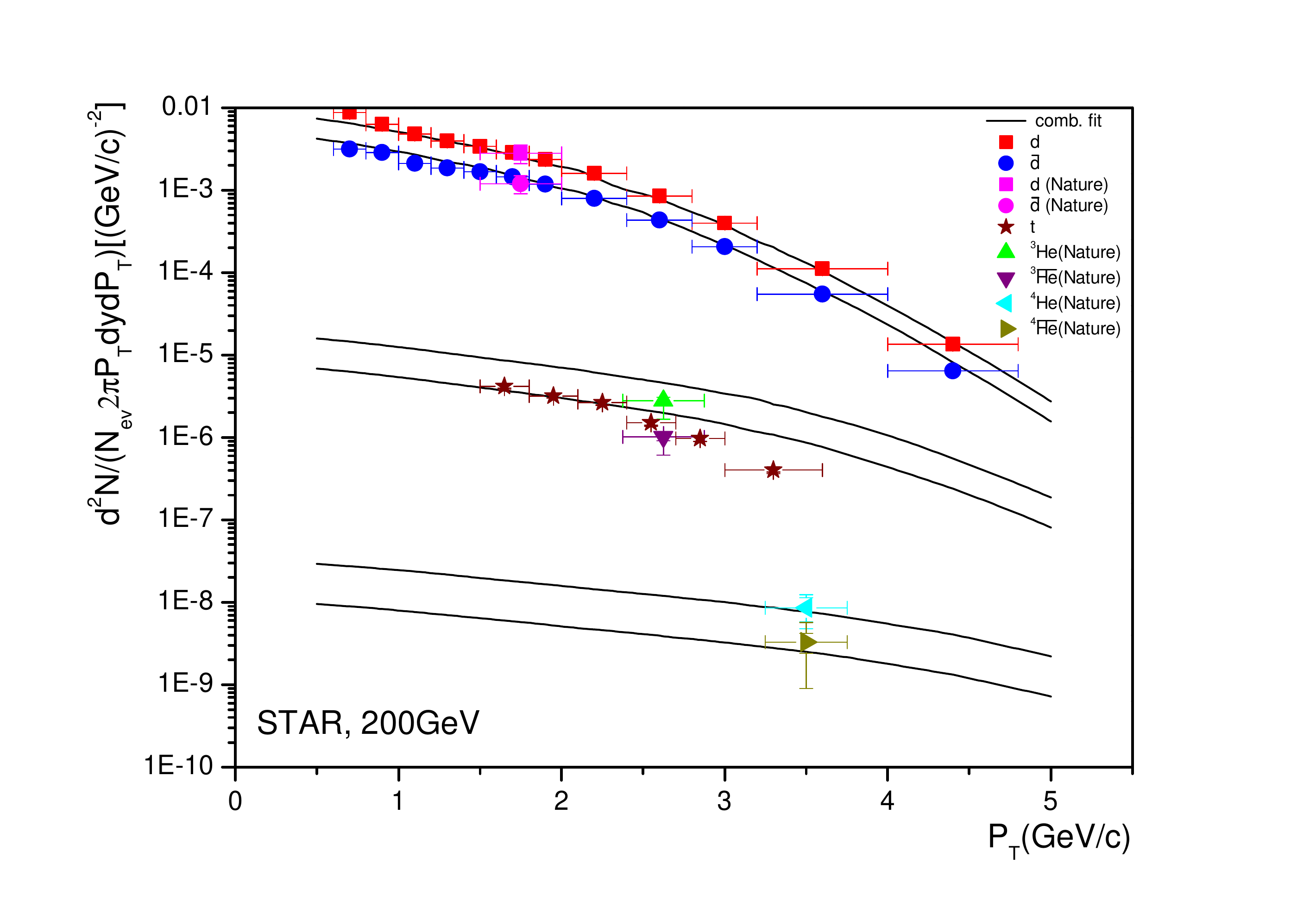}
\end{minipage}

\vspace{-0.5cm}

\begin{minipage}{0.5\textwidth}
\includegraphics[width=0.95\textwidth]{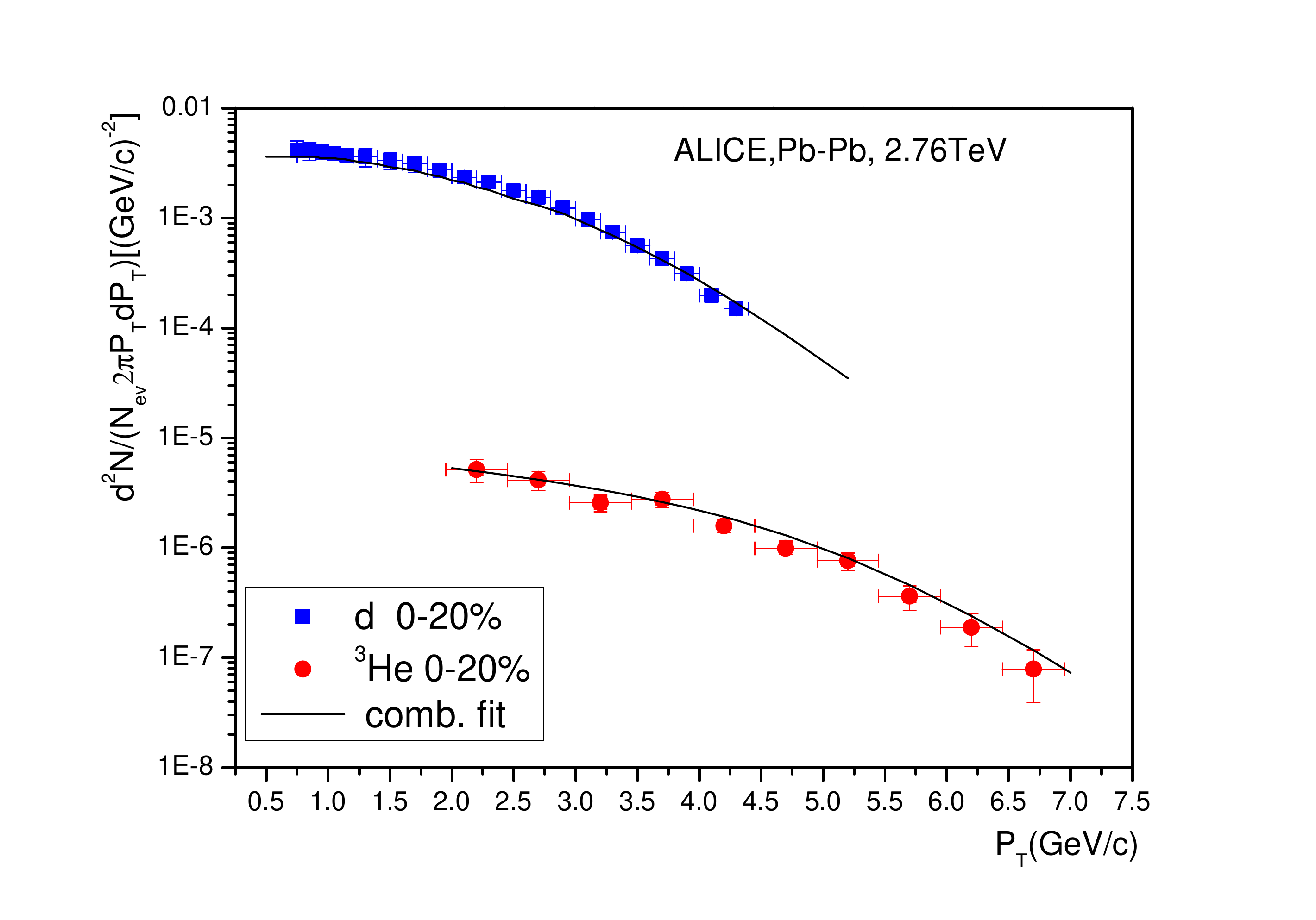}
\end{minipage}

\caption{Transverse-momentum spectra of light anti-/nuclei at RHIC (upper panel) and the
LHC (lower panel) in a common blastwave description at kinetic freezeout, compared 
to STAR data~\cite{Abelev:2009ae,Agakishiev:2011ib,Yu:2017bxv,Liu:2018qm} for 
0-10\,\% central Au+Au(0.2\,TeV) collisions and ALICE data~\cite{Adam:2015vda} 
for 0-20\,\% central Pb+Pb(2.76\,TeV) collsions.
}
\label{fig_pt}
\efig

\section{Conclusions}
\label{sec:concl}
We have revisited the problem of the production of light nuclei and antinuclei in high-energy
heavy-ion collisions. This was triggered by the surprising finding that their production 
yields follow chemical-freezeout systematics while their small binding energies of a few MeV
per nucleon suggest that these states should not exist at pertinent temperatures of 150-160\,MeV.
We have found that, employing the concept of chemical freezeout in the subsequent hadronic
evolution to kinetic freezeout, as previously implemented in different contexts, preserves the 
ratios and yields of anti-/nuclei evaluated at chemical freezeout down to thermal freezeout (with 
the exception of A=3 nuclei at RHIC, at both freezeouts). We also found the momentum spectra of 
the anti-/nuclei at kinetic freezeout to be compatible with a common decoupling with flow velocities 
extracted from light-hadron spectra. While this enables the formation of the nuclei to occur 
at much lower temperatures of $T_{\rm fo}$$\simeq$100\,MeV, one can still argue that this is 
too high for them to exist. However, both the momentum 
distributions and yields of hadrons are ``frozen" at this point and thus should allow for the 
formation of nuclear bound states even in the aftermath of the hadronic kinetic freezeout, which 
still conserves the relative chemistry, much like in the evolution from chemical to kinetic 
freezeout, but without further ``cooling" and associated changes in momentum distributions. 
In other words, the chemical reactions for the formation of anti/-nuclei (which are expected
to have large cross sections) could continue to still lower densities, to a ``nuclear freezeout", 
without significantly affecting the momentum distributions nor the hadronic chemistry of the 
hadronic bulk system. This picture appears to be 
closely related to coalescence mechanisms, but the evaluations are carried out without additional
coalescence parameters. After all, coalescence models should, in principle, encode the correct
equilibrium limit, an issue that is also pertinent to the hadron formation processes from an 
equilibrated QGP as well as in the kinetic regime of intermediate transverse 
momenta~\cite{Ravagli:2007xx}.   
We therefore believe that our results can contribute to a better understanding of the
formation of light nuclei in URHICs.

\vspace{0.5cm}

\noindent 
{\bf Acknowledgment}\\
We thank Zhenyu Ye and Anton Andronic for help with the STAR and ALICE data, respectively.
Some of the ideas of this work have been triggered by discussions at the Terzolas meeting 
on ``Heavy-Ion Physics in the 2020's" (Terzolas, Italy, May 19-21, 2018). 
This work has been supported by the U.S. National Science Foundation under grant no.~PHY-1614484.
\\

\noindent
{\bf Note added}\\
Shortly before submission of the present ms., a related work appeared~\cite{Oliinychenko:2018ugs} 
discussing the problem of anti-/deuteron production at the LHC using a hadronic transport approach.

\end{document}